\documentclass[aps,prc,twocolumn,superscriptaddress]{revtex4-1}

\usepackage{amsmath,amssymb,graphicx}
\usepackage{url}
\usepackage[colorlinks,urlcolor=blue]{hyperref}

\begin{document}

\title{Rapidity asymmetry of jet-hadron correlation as a robust signal of diffusion wake induced by di-jets in high-energy heavy-ion collisions}

\author{Zhong Yang}
\email[]{zhongyang94@ccnu.edu.cn}
\affiliation{Key Laboratory of Quark and Lepton Physics (MOE) \& Institute of Particle Physics, Central China Normal University, Wuhan 430079, China}
\affiliation{Department of Physics and Astronomy, Vanderbilt University, Nashville, TN}

\author{Xin-Nian Wang}
\email[]{xnwang@ccnu.edu.cn}
\affiliation{Key Laboratory of Quark and Lepton Physics (MOE) \& Institute of Particle Physics, Central China Normal University, Wuhan 430079, China}

\begin{abstract}
Diffusion wake accompanying a Mach cone is a unique feature of the medium response to  projectiles traveling at a speed faster than the velocity of sound. This is also the case for  jet-medium interaction inside the quark-gluon plasma in high-energy heavy-ion collisions. It leads to a depletion of soft hadrons in the opposite direction of the propagating jet and has been recently observed in $Z$-jet events of Pb+Pb collisions at LHC. In di-jet events, however, the diffusion wake of one jet usually overlaps with the medium-induced hadron enhancement of other jet without a clear signal except a reduction of the hadron enhancement, unless there is a large rapidity gap between the two jets. We propose to use the rapidity asymmetry of jet-hadron correlations in di-jets with a finite rapidity gap relative to that without,  as a robust and background-free signal of the diffusion wake. The asymmetry emerges because the diffusion wake of one jet is shifted to a finite rapidity relative to the other jet. Consequently, a depletion of soft hadrons appears in the shifted rapidity region of the diffusion wake and an enhancement in the rapidity region of the other jet whose soft hadron enhancement is no longer or less reduced by the diffusion wake as in di-jets without a rapidity gap.  We predict the rapidity asymmetry using both theoretical and mixed-event background subtraction for different values of the rapidity gap within the CoLBT-hydro model.  Future measurements of this rapidity asymmetry with high statistics data on di-jets should provide more precise insights into the jet-induced diffusion wake and properties of the quark-gluon plasma. 

\end{abstract}

\maketitle

\noindent 1. {\bf Introduction:}
Mach cones, such as the sonic boom of a supersonic jet fighter, are superpositions of Mach waves generated by a projectile that travels at a speed faster than the velocity of sound \cite{mach}. These also occur in high-energy heavy-ion (A+A) collisions, where jet particles propagating at the velocity of light interact with the QCD medium and deposits a fraction of their energies and momenta into the quark-gluon plasma (QGP)~\cite{Bjorken:1982tu, Thoma:1990fm, Braaten:1991we, Gyulassy:1993hr, Baier:1996kr, Zakharov:1996fv, Gyulassy:1999zd, Wiedemann:2000za, Wang:2001ifa, Arnold:2002ja, Djordjevic:2006tw, Qin:2007rn}. Such energy loss induces a medium response that manifests as Mach-cone-like excitations on the femtometer scale~\cite{Casalderrey-Solana:2004fdk, Stoecker:2004qu, Ruppert:2005uz, Gubser:2007ga, Neufeld:2008fi,Qin:2009uh, Li:2010ts, Bouras:2012mh, Ayala:2016pvm, Yan:2017rku, Casalderrey-Solana:2020rsj}. Investigating the jet-induced medium response can provide valuable insights into the properties of the QGP, as the creation and evolution of the medium response depend on the dynamical and transport properties of the QGP medium.

Many theoretical and experimental studies of jet-induced medium response in the past have focused on the search for signals of the jet-induced Mach cone~\cite{Cao:2020wlm, Betz:2010qh, Ma:2010dv, Tachibana:2014lja, Casalderrey-Solana:2016jvj, Tachibana:2017syd, Chen:2017zte, He:2018xjv, Zhang:2018urd, Luo:2018pto, Pablos:2019ngg, Chen:2020tbl, Tachibana:2020mtb, Yang:2022yfr,Du:2022oaw, Mehtar-Tani:2022zwf, Yang:2023dwc, JETSCAPE:2023hqn, Xiao:2024ffk, Li:2024pfi, Feng:2024tmc, Bossi:2024qho, Barata:2024ieg, CMS:2018jco,CMS:2018mqn, ATLAS:2020wmg, CMS:2021otx,  PHENIX:2024twd, Arslandok:2023utm, Chakraborty:2006md}. Unfortunately, medium-induced soft gluon radiation of a propagating jet also contributes to particle production in the jet direction which cannot be uniquely separated from soft hadrons due to jet-induced Mach cone. Jet-induced diffusion wake~\cite{Gubser:2007ni,Betz:2008ka,Yang:2021qtl, Yang:2022nei,ATLAS:2024prm, CMS:2024fli}, on the other hand, is a unique aspect of the jet-induced medium response accompanying the Mach cone that depletes soft hadrons from the bulk in the opposite direction of jet. 
Such depletion of soft hadrons due to jet-induced diffusion wake in A+A collisions with respect to p+p can be measured through the azimuthal dependence of jet-hadron correlations \cite{Chen:2017zte, Yang:2021qtl,CMS:2021otx,yjt}. The challenge in searching for such a depletion of soft hadrons in the opposite direction of a jet in A+A collisions is the subtraction of background which is actually enhanced
due to medium modification of mini-jets from initial multiple parton interactions (MPI) \cite{Yang:2021qtl,CMS:2021otx}. Though one can use the technique of recombinatory background subtraction to overcome the complication of the medium modification of MPI \cite{Yang:2021qtl},

the diffusion wake can be directly observed in two-dimensional jet-hadron correlations in azimuthal angle and rapidity.
As shown in a recent study \cite{Yang:2022nei}, a clear diffusion wake valley in rapidity on top of the MPI ridge along the azimuthal angle in jet-hadron correlation in $\gamma/Z$-jet events  of Pb+Pb collisions at the LHC energy offers a direct  identification of the signal of jet-induced diffusion wake. Such a diffusion wake valley in $Z$-hadron correlation was indeed observed recently by CMS \cite{yjt}, providing direct evidence of jet-induced medium response in the QGP in Pb+Pb collisions at the LHC. A similar observation has been made by the ATLAS in $\gamma$-jet events \cite{ATLAS:2023fjw,ATLAS:2024prm}. Current experimental data from CMS and ATLAS, however,  still exhibit relatively large statistical errors due to limited number of Z/$\gamma$-jet events in A+A collisions at LHC. It is therefore crucial to investigate the signal of jet-induced diffusion wake in di-jet events, which are far more abundant at LHC.
 
\begin{figure}[h!]
\centering
    \includegraphics[width=0.48\textwidth]{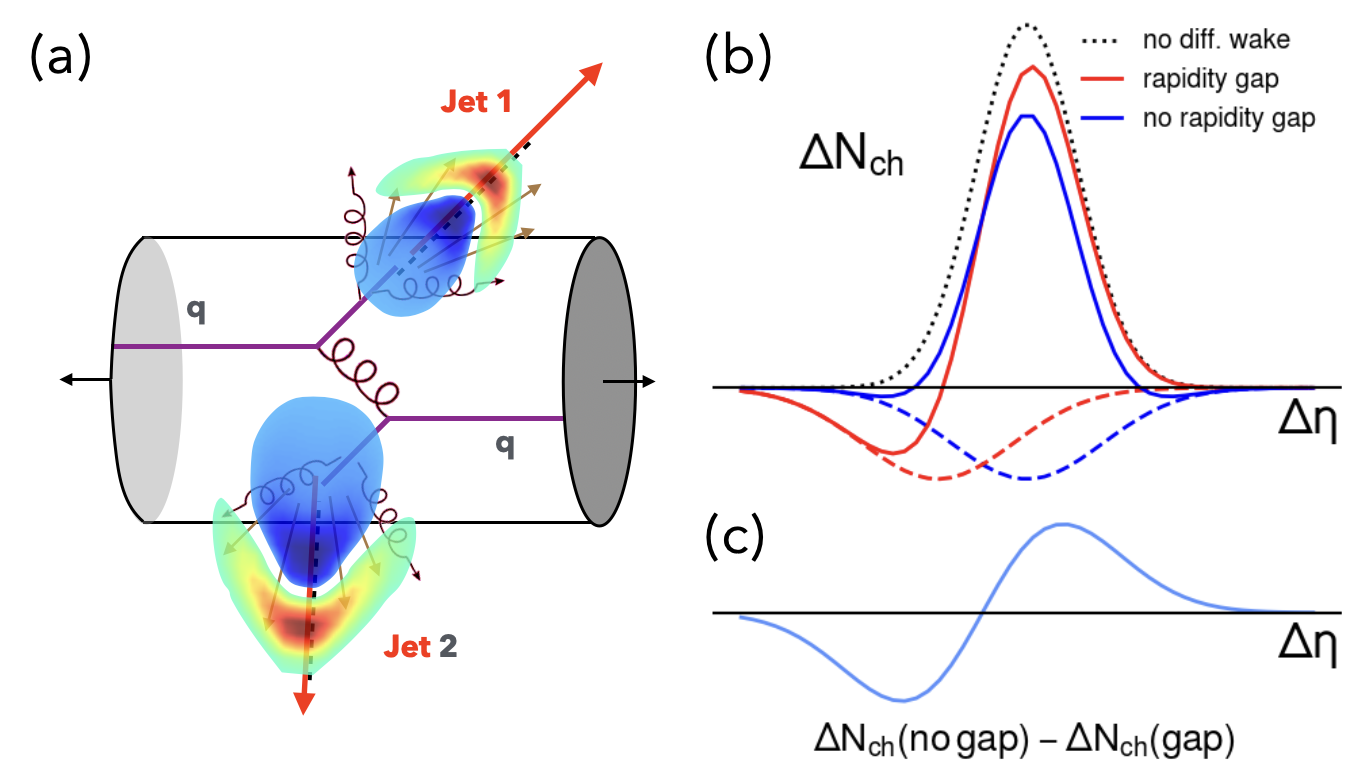}
	\caption{(a) Wake front (red-yellow-green), diffusion wake (blue) and induced gluon radiation in di-jet events in heavy-ion collisions. (b) Illustration of the effect of the diffusion wake (dashed) from one jet on the jet-hadron correlation in the other jet with (red) and without (blue) a rapidity gap and (c) the rapidity asymmetry of their difference.}
	\label{illus}
\end{figure}

For a di-jet within the same rapidity range, the diffusion wake induced by one jet in the opposite azimuthal angle overlaps completely with the soft hadron enhancement from the wake front and induced gluon radiation of other jet as illustrated in Fig.~\ref{illus}. The diffusion wake in this case reduces the medium-induced hadron enhancement but leaves no other clear signals as illustrated in (blue curves) Fig.~\ref{illus} (b).  One can avoid such a situation by selecting di-jets with a finite rapidity gap as first suggested in Ref.~\cite{Pablos:2019ngg}.  In this Letter we propose to use the rapidity asymmetry of jet-hadron correlations in di-jets with a rapidity gap as the signal of jet-induced diffusion wake which is robust and background-free even for di-jets within a moderate rapidity coverage in high-energy A+A collisions. Using CoLBT-hydro model \cite{Chen:2017zte,Chen:2020tbl,Zhao:2021vmu}, we show this asymmetry arises from the depletion of soft hadrons in the rapidity region of the diffusion wake and an enhancement in the rapidity region of the reference jet whose soft hadrons are no longer or less reduced by the diffusion wake, relative to di-jets without a rapidity gap, as illustrated in Figs.~\ref{illus} (b) and (c).  We will also predict the dependence of the asymmetry on the rapidity gap.

\noindent 2. {\bf Imaging the diffusion wake in di-jet events:} 
In Z/$\gamma$-jet events, the signal of a diffusion wake is distinct in the Z/$\gamma$ direction \cite{Yang:2022nei, yjt}  since Z/$\gamma$ does not interact with the QGP medium and induce any medium modifications in its direction.  In di-jet events, however, as illustrated in Fig.~\ref{illus} (a), both jets undergo energy loss, inducing medium response and gluon radiation, leading to enhancement of soft hadrons in their directions. Since the two jets are essentially back-to-back in azimuthal angle, the diffusion wake induced by one jet overlaps with and is therefore overwhelmed by the wave front and induced gluon radiation generated by the other jet,
making it impossible to observe the depletion of soft hadrons as the signal of the diffusion wake of either of the two jets in the azimuthal angle distributions.

Because of the 3D nature of jet structure and medium response, we have proposed to use jet-hadron correlation in both azimuthal angle and rapidity to look for the signal of the diffusion wake. A valley in rapidity on top of the MPI ridge along the opposite direction of the jet in azimuthal angle was found \cite{Yang:2022nei} and indeed observed by CMS \cite{yjt} in $Z$-jet events of Pb+Pb collisions at LHC. One can do the same for di-jet events.

To calculate jet-hadron correlations in di-jet events in Pb+Pb collisions, we employ the CoLBT-hydro model~\cite{Chen:2020tbl, Chen:2017zte, Zhao:2021vmu} to simulate di-jet transport in the QGP medium with the initial di-jet configurations generated by PYTHIA8~\cite{Sjostrand:2007gs, Sjostrand:2014zea, Bierlich:2022pfr}. The CoLBT-hydro model integrates the Linear Boltzmann Transport (LBT) model~\cite{He:2015pra, Luo:2023nsi, Cao:2016gvr, Cao:2017hhk, Wang:2013cia, Xing:2019xae} with the event-by-event (3+1)D CCNU-LBNL viscous hydrodynamic (CLVisc) model~\cite{Pang:2012he, Pang:2014ipa, Pang:2018zzo, Wu:2021fjf} to simulate the concurrent evolution of parton showers and the QGP medium. 
For more detailed description of the CoLBT-hydro model see Refs.~\cite{Chen:2020tbl, Chen:2017zte, Zhao:2021vmu}.
The final hadron spectra associated with jets are calculated after the subtraction of background from the same hydro event without the di-jet. We refer this as the theoretical background subtraction.

\begin{figure}[h!]
\centering
    \includegraphics[width=0.48\textwidth]{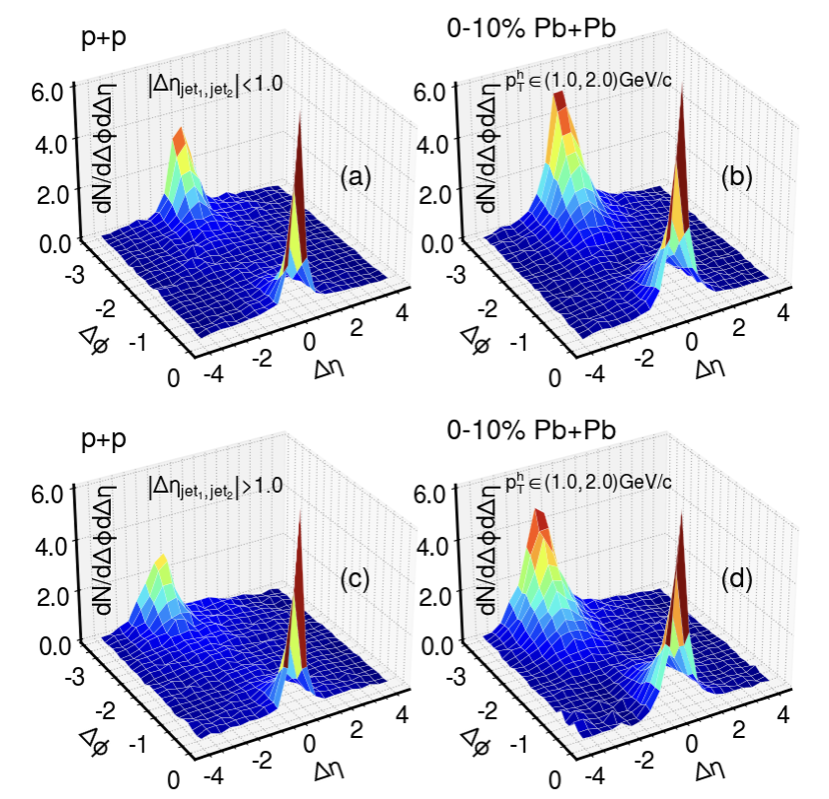}
	\caption{Correlations between leading full jets ($p_T^{\rm jet_1}> 120$ GeV/$c$) and soft charged hadrons ($1 \le p_T^h\le 2$ GeV/$c$ ) in p+p (a,c) and 0-10\% central Pb+Pb (b,d) collisions at $\sqrt{s_{\rm NN}}=5.2$ TeV as a function of $\Delta\eta=\eta_h-\eta_{{\rm jet}_1}$ and $\Delta\phi=\phi_h-\phi_{{\rm jet}_1}$ for two values of the di-jet rapidity gap $|\Delta\eta_{\rm jet_1,jet_2}|<1$ (a,b) and $|\Delta\eta_{\rm jet_1,jet_2}|>1$ (c,d).  The sub-leading jets are required to have $p_T^{\rm jet_2}>90$ GeV/$c$ and the azimuthal angle difference $|\Delta\phi_{\rm jet_1,jet_2}|>\pi/2$. The jet cone size is $R=0.4$.}
	\label{corr3D}
\end{figure}

For this study, we simulate jet production in p+p and central 0-10\% Pb+Pb collisions at $\sqrt{s_{\rm NN}}=5.02$ TeV within the CoLBT-hydro model for a large range of hard scale $\hat p_T$ and reconstruct full jets using the FASTJET anti-kT algorithm \cite{Cacciari:2011ma} from the hadron spectra after the theoretical background subtraction. We then calculate correlations between leading full jets and charged hadrons in di-jet events with $p_T^{\rm jet_1}> 120$ GeV/$c$  for the leading jet and $p_T^{\rm jet_2}>90$ GeV/$c$ for the sub-leading jet as a function of $\Delta\eta=\eta_h-\eta_{{\rm jet}_1}$ and $\Delta\phi=\phi_h-\phi_{{\rm jet}_1}$ as shown in Fig.~\ref{corr3D}, for two values of the rapidity gap $|\Delta\eta_{\rm jet_1,jet_2}|\equiv|\eta_{{\rm jet}_1}-\eta_{{\rm jet}_2}|<1$ (upper panels) and $|\Delta\eta_{\rm jet_1,jet_2}|>1$ (lower panels). The jet cone size is R=0.4 and only jets with $\left | \eta_{\rm{jet_1},\rm{jet_2}}\right |< 1.6$ and $|\Delta\phi_{\rm jet_1,jet_2}|\equiv|\phi_{{\rm jet}_1}-\phi_{{\rm jet}_2} |>\pi/2$ are considered. The jet-hadron correlations are normalized by the number of jet pairs. We choose the sign of the rapidity in each event such that $\eta_{\rm jet_1}>\eta_{\rm jet_2}$. This is crucial for our study of the rapidity asymmetry caused by the diffusion wake in jet-hadron correlations.

In general, there are two peaks in the jet-hadron correlation. One at $\Delta\phi=0$ is the near-side correlation for hadrons from the leading jet and another one at $\Delta\phi=\pi$ is the away-side correlation for hadrons from the sub-leading jet. Given our convention of jet rapidity ordering, the away-side peak and the associated diffusion wake of the sub-leading jet are always in the negative region of the rapidity $\Delta\eta$. We see both the near-side and away-side correlations for soft hadrons are enhanced and broadened in central Pb+Pb collisions as compared to p+p, due to soft hadrons from medium-induced gluon radiation and jet-induced medium response which includes both the wake front of one jet and diffusion wake from the other. The correlations for high $p_{T}$ hadrons should be suppressed due to parton energy loss. The broadening and enhancement are stronger in away-side correlations than the near-side, since sub-leading jets traverse longer distance and lose more energy than the leading jets due to trigger bias.

For small values of the di-jet rapidity gap $|\Delta\eta_{\rm jet_1,jet_2}|<1$, the peaks of near-side and away-side correlation are close to each other at around $\Delta\eta=0$ as shown Figs.~\ref{corr3D} (a) and (b).
Since the diffusion wake is in the same rapidity region but opposite in azimuthal angle of the jet,  one cannot see any apparent effects of the diffusion wakes induced by either of the two jets when their rapidity gap is small, as we expect. The broadening and enhancement in the jet-hadron correlation of both jets in A+A collisions, however,  should be reduced by the diffusion wakes of the other jets.

For dijets with a large rapidity gap, the away-side peak of the jet-hadron correlation from the sub-leading jets is at finite rapidity as we see in Figs.~\ref{corr3D} (c) and (d). The corresponding diffusion wake of the sub-leading jet will also be at a large rapidity gap away from the leading jet. The depletion of soft hadrons due to this diffusion wake with a rapidity shift away from the leading jet becomes clearly visible as a dip in the near-side correlation. The diffusion wake of the leading jet cannot be seen clearly in the direction of the sub-leading jet due to the broader and enhanced distribution of the away-side jet-hadron correlation. It, however, should still influence the away-side correlation and causes a rapidity asymmetry as we show next.

\noindent 3. {\bf Rapidity asymmetry of jet-hadron correlation:} 
Given the above 3D structure of the jet-hadron correlations, we can project the correlations onto the rapidity in both near ($\Delta\phi<\pi/2$) and away-side ($\Delta\phi>\pi/2$) region of the leading jet and focus on the medium modifications of the projected correlation in rapidity $\Delta \eta=\eta_h-\eta_{\rm jet_1}$, which is defined as the difference between jet-hadron correlation in Pb+Pb and p+p collisions,
\begin{equation}
    \Delta N_{\rm AA}=\int d\Delta\phi\left[\frac{dN_{\rm AA}}{d\Delta\phi d\Delta\eta}-\frac{dN_{\rm pp}}{d\Delta\phi d\Delta\eta}\right].
\end{equation}

\begin{figure}[h!]
\centering
	\includegraphics[width=0.48\textwidth]{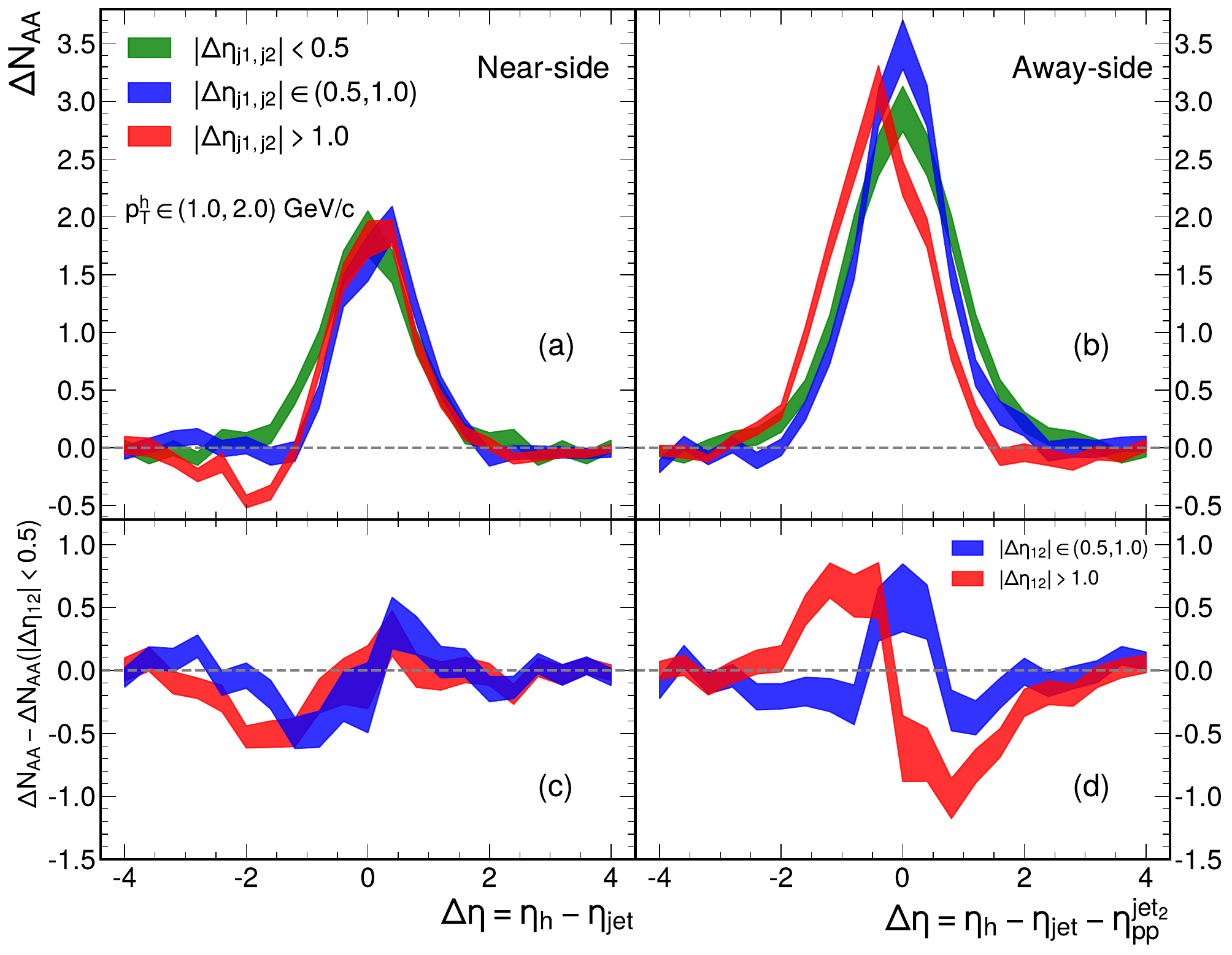}
	\caption{(a) Medium modification $\Delta N_{AA}$ of near-side and (b) away-side jet-hadron rapidity correlation in di-jets with three different ranges of rapidity gap $|\Delta\eta_{\rm jet_1,jet_2}|$ in 0-10\% central Pb+Pb collisions  and (c,d) the corresponding asymmetry $\Delta N_{AA}-\Delta N_{AA}(|\Delta\eta_{\rm jet_1,jet_2}|<0.5)$.}
	\label{rap-distr}
\end{figure}

Shown in Figs.~\ref{rap-distr} (a) and (b) are the medium modifications of the near and away-side jet-hadron correlations, respectively, in 0-10\% central Pb+Pb collisions at $\sqrt{s_{\rm NN}}=5.02$ TeV for di-jets with three different values of the rapidity gap $|\Delta\eta_{\rm jet_1,jet_2}|<0.5$, $0.5<|\Delta\eta_{\rm jet_1,jet_2}|<1$ and $|\Delta\eta_{\rm jet_1,jet_2}|>1$. For the away-side correlations, we have shifted the medium modification distributions by the rapidity of the away-side peaks in p+p collisions for each given di-jet rapidity gap $|\Delta\eta_{\rm jet_1,jet_2}|$. We further symmetrize the modification distributions with the smallest gap $|\Delta\eta_{\rm jet_1,jet_2}|<0.5$ with respect to their peaks as the references for calculating the rapidity asymmetry of jet-hadron correlations in di-jets with large rapidity gaps.

As we see in the jet-hadron correlations (both near and away-side) in di-jets with a small rapidity gap $|\Delta\eta_{\rm jet_1,jet_2}|<0.5$ (green curves), the diffusion wake of one jet (jet 1) overlaps with the hadron enhancement from the wake front and soft gluon radiation in the other jet (jet 2). One therefore cannot see any apparent effect of the diffusion wake of jet 1 on the jet-hadron correlation in jet 2, except that it reduces the enhancement of soft hadrons in jet 2. This is consistent with the illustration in Fig.~\ref{illus} (b). The net enhancement in the sub-leading jet (away-side) is larger than the leading jet (near-side)  because of trigger bias and larger energy loss of the sub-leading jet.

For di-jets with large rapidity gaps, $0.5<|\Delta\eta_{\rm jet_1,jet_2}|<1$ and $|\Delta\eta_{\rm jet_1,jet_2}|>1$, the diffusion wake of the sub-leading jet does not coincide with the medium-induced enhancement of the leading jet and vice versa. It leads to a net depletion (a negative dip) of the near-side jet-hadron correlation in the rapidity region of the sub-leading jet or its diffusion wake, as we can see in Fig.~\ref{rap-distr} (a) (blue and red curves).  This gives rise to an asymmetric medium-modified jet-hadron correlation in rapidity.  Because of trigger bias, the energy loss of the sub-leading jet is larger, so is the soft hadron enhancement. As a consequence, the diffusion wake of the leading jet is not big enough to result in a sizable net depletion of hadrons, but just a reduction in the away-side jet hadron correlation. This reduction also leads to an asymmetric medium-modified  away-side jet-hadron correlation in  rapidity as we see in Fig.~\ref{rap-distr} (b).

To characterize the asymmetric rapidity jet-hadron correlations caused by diffusion wakes in di-jet events, we plot in Figs.~\ref{rap-distr} (c) and (d) the rapidity asymmetry defined as the differences between medium modifications of near and away-side jet-hadron correlation, respectively, in di-jets with large and small rapidity gaps $\Delta N_{\rm AA}-\Delta N_{\rm AA}(|\Delta\eta_{\rm jet_1,jet_2}|<0.5)$. 
We want to emphasize that this rapidity asymmetry is background-free. If one fixes the rapidity of the leading (trigger) jet and varies the rapidity of the sub-leading jets, the backgrounds from both the bulk hadron production and MPI are the same for di-jet events with different rapidity gaps which will be completely canceled in the rapidity asymmetry as we define here. Consequently, it can be measured directly in experiments without any need of background subtraction.
As one can see, the diffusion wake (depletion of hadrons) in di-jets with a large rapidity gap is shifted in rapidity by the amount of the gap relative to the reference jet as compared to the jet-hadron correlation with no or small di-jet rapidity gap (symmetrized). As we have illustrated in Figs.~\ref{illus} (b) and (c), this shift causes the rapidity asymmetry as seen in Figs.~\ref{rap-distr} (c) and (d). It leads to a depletion of soft hadrons (negative asymmetry) in the rapidity region of the diffusion wake and an enhancement (positive asymmetry) in the rapidity region of the reference jet whose soft hadron enhancement is no longer or less reduced by the diffusion wake as compared to di-jets without a rapidity gap. The peak positions of the negative and positive asymmetry increases with the di-jet rapidity gap.  Note that the amplitude of the rapidity asymmetry in away-side correlation is bigger than the near-side since sub-leading jet and the diffusion wake of the leading jet traverse a longer distance.  The correlations for jets at finite rapidity should also be influenced by the longitudinal flow and gradient. Since the jet rapidity ordering we use is independent of the longitudinal direction, only the widths of the correlations (both the enhancement and the diffusion wake) are influenced by the flow and gradient, but not peak (dip) positions.

\begin{figure}[h!]
\centering
	\includegraphics[width=0.42\textwidth]{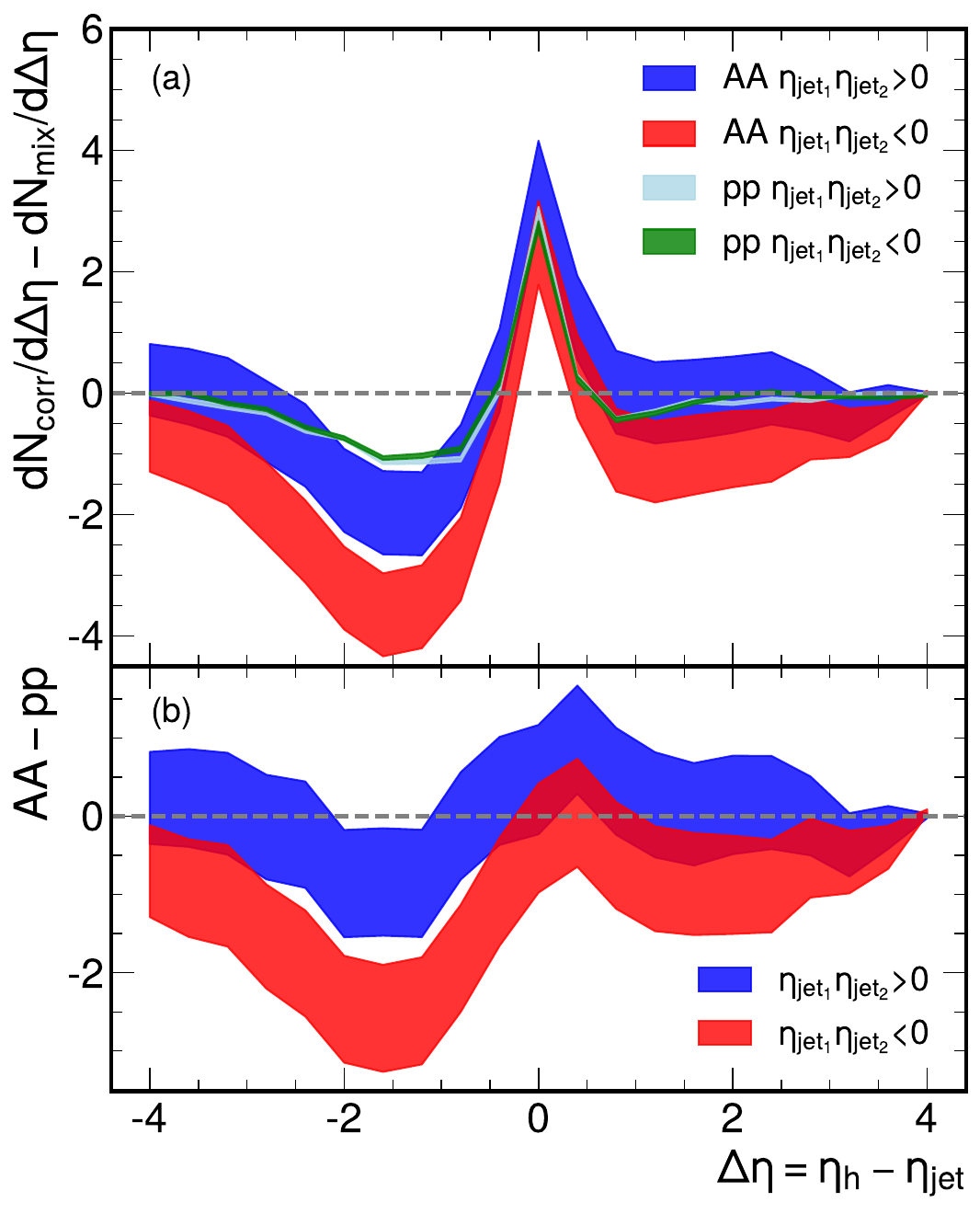}
	\caption{(a) Near-side jet-hadron correlations for same-hemisphere (blue, cyan) and opposite-hemisphere (red, green) di-jets after mix-event background subtraction in p+p and 0-10\% central Pb+Pb collisions at $\sqrt{s_{\rm NN}}=5.02$ TeV and (b) the medium modification.}
	\label{distr-mix}
\end{figure}

Instead of the subtraction of theoretical background in our model calculations, mixed events are usually used for background subtraction in experimental analyses \cite{yjt}. Following the same technique, we calculate the correlation between the trigger jet and all charged hadrons, including the hydro background, in the same event. The true correlations are obtained after subtraction of the correlation between the trigger jet in one event and all charged hadrons from another event with the same selection criteria. Shown in Fig.~\ref{distr-mix} are the near-side jet-hadron correlations in p+p and 0-10 \% central Pb+Pb collisions (upper panel) and their difference (lower panel). Considering the limited range of rapidity coverage in most experiments at RHIC and LHC, we divide di-jet events into two sub-events according to the relative polar angle of the two jets: same-hemisphere ($\eta_{{\rm jet}_1}\eta_{{\rm jet}_2}>0$) and opposite-hemisphere (($\eta_{{\rm jet}_1}\eta_{{\rm jet}_2}<0$). On average, the opposite-hemisphere di-jets have a larger rapidity gap than the same-hemisphere. Event mixing is done among events in the same sub-events. Again, we use the convention of rapidity ordering of leading and sub-leading jet, $\eta_{\rm jet_1}>\eta_{\rm jet_2}$ in all events. The jet-hadron correlations after subtraction of the mixed-event background are asymmetrical (upper panel) in both p+p and Pb+Pb collisions. The asymmetry in p+p, which is independent of the di-jet rapidity gap, is caused by the interference between hadron production in jets and the beam remnants. The asymmetry in Pb+Pb collisions is caused by the diffusion wake which increases with the rapidity gap. The medium modification of such jet-hadron correlations, defined as the difference between A+A and p+p collisions, resembles the rapidity asymmetry as we have presented earlier in Figs.~\ref{rap-distr}. One can clearly see the depletion of soft hadrons in the near-side jet-hadron correlation due to the diffusion wake of the sub-leading jet. The enhancement in the positive rapidity region  is due to reduced effect of the diffusion wake in this region relative to that in the mix events. One can also see the asymmetry still exists, though is weaker, for same-hemisphere di-jets. A depletion of jet-hadron correlation at large $\Delta\eta$ in single inclusive jets is also expected with this mix-event background subtraction, though the signal is much weaker.

\noindent 4. {\bf Summary:}
In this Letter, we propose a novel method to identify the signal of diffusion wake induced by di-jets with finite rapidity gap in high-energy heavy-ion collisions. Using the CoLBT-hydro model simulations, we show jet-hadron correlations in di-jets with a small or no rapidity gap do not show any apparent signals of the diffusion wake. The jet-hadron correlations in di-jets with a large rapidity gap, however, show a significant depletion of soft hadrons in the rapidity region of the diffusion wake at a rapidity distance away from the reference jet. This rapidity separation between the diffusion wake and the reference jet leads to a rapidity asymmetry of jet-hadron correlations in di-jets with a large rapidity gap relative to that with no or a small rapidity gap.  It shows a depletion of soft hadrons in the rapidity region of the diffusion wake and an enhancement in the rapidity region of the reference jet where medium-induced enhancement is less reduced by the shifted  diffusion wake. Most importantly, this rapidity asymmetry is background free and significantly simplifies experimental measurements.
We also show the same feature exists in the jet-hadron correlations with the mix-event background subtraction. With higher statistics data on di-jets in the current and future experiments at RHIC and LHC, this study provides an effective method for more detailed experimental study of the jet-induced diffusion wake in high-energy heavy-ion collisions.

\noindent {\bf Acknowledgement:} We thank Daniel Pablos for suggesting this study, Luna Chen and Rithya Kunnawalkam Elayavalli for helpful discussions. This work is supported in part by the China Postdoctoral Science Foundation under Grant No. 2024M751059 and No. BX20240134(ZY), by NSFC under Grant No. 1193507 and by the Guangdong MPBAR with No. 2020B0301030008. Computations in this study are performed at the NSC3/CCNU and NERSC under the award NP-ERCAP0032607. 

\noindent {\bf Data Availability:} The processed data that support the findings of this Letter are openly available in \cite{yang_2025_16290143}. The full raw dataset is available from the corresponding author upon reasonable request for validation of the findings presented in this article.
\bibliography{Refsdw}

\end{document}